\documentclass[twocolumn,showpacs,preprintnumbers,amsmath,amssymb]{revtex4}

\usepackage{graphicx}
\usepackage{dcolumn}
\usepackage{bm}

\begin{document}

\title{Momentum entanglement and disentanglement between atom and photon}

\author{Rui Guo}
\author{Hong Guo}\thanks{Author to whom correspondence should be
addressed. E-mail: hongguo@pku.edu.cn, phone: +86-10-6275-7035, Fax:
+86-10-6275-3208.}

\affiliation{CREAM Group, School of Electronics Engineering $\&$
Computer Science, Peking University, Beijing
100871, P. R. China\\}%

\date{\today}

\begin{abstract}
With the quantum interference between two transition pathways, we
demonstrate a novel scheme to coherently control the momentum
entanglement between a single atom and a single photon. The
unavoidable disentanglement is also studied from the first
principle, which indicates that the stably entangled atom--photon
system with superhigh degree of entanglement may be realized with
this scheme under certain conditions.
\end{abstract}

\pacs{03.65.Ud, 42.50.Vk, 32.80.Lg }.

\maketitle

\section{Introduction}

In recent years, entanglement with continuous variables attracts
substantial attention for its importance in quantum nonlocality
\cite{EPR} and quantum information processing (QIP) \cite{rmp}. As
a physical realization, momentum entanglement has been extensively
studied both theoretically \cite{3-D spontaneous, GR,
photoionization, scattering, Singlephoton} and experimentally
\cite{exp}. With momentum entanglement between atom and photon, it
is possible to define the best localized single--photon wavepacket
even in free space \cite{Singlephoton}, and realize the highest
degree of continuous entanglement \cite{scattering} up to date.

As known, photon emitted from atom will recoil and be entangled
with the atom \cite{3-D spontaneous,GR,scattering} due to momentum
conservation. In order to coherently manipulate the entanglement,
in this paper, we propose a novel scheme to control the
entanglement through the atomic spontaneously generated coherence
(SGC). With the configuration in Fig. 1(a), we find that, the
recoil--induced--entanglement will be affected by the interference
between different transitions from the atomic upper levels, and
can be effectively controlled by an auxiliary coupling field if
the dipoles for the transitions are parallel. This scheme,
compared with the Raman scattering \cite{scattering} and resonant
scattering \cite{GR}, could be more efficient in producing
superhigh degree of entanglement, since the controlling light is
classical and need not be far detuned.

To consider the scheme in a more realistic situation, we study the
unavoidable process of disentanglement following the generation of
entanglement. From the first principle, we obtain the master
equation and the characteristic time scale for the
disentanglement. Comparing the two processes, we yield an upper
bound for the degree of entanglement that may be steadily produced
with the scheme. Under realistic conditions \cite{SGC exp}, it is
shown that the robust atom--photon entangled pair can be produced
with superhigh degree of entanglement.

\section{entanglement generation by interference}
\begin{figure}
\centering
\includegraphics[height=3.5cm]{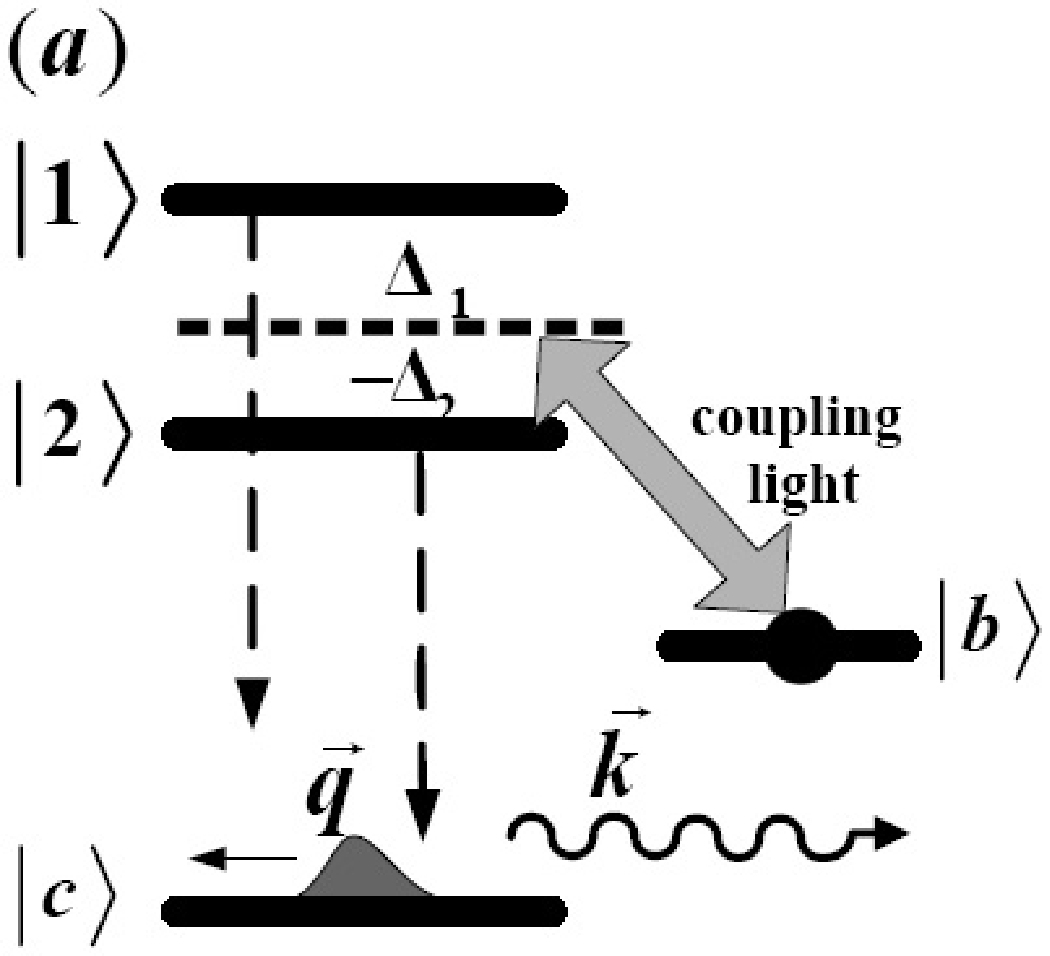}\ \ \ \includegraphics[height=3.5cm]{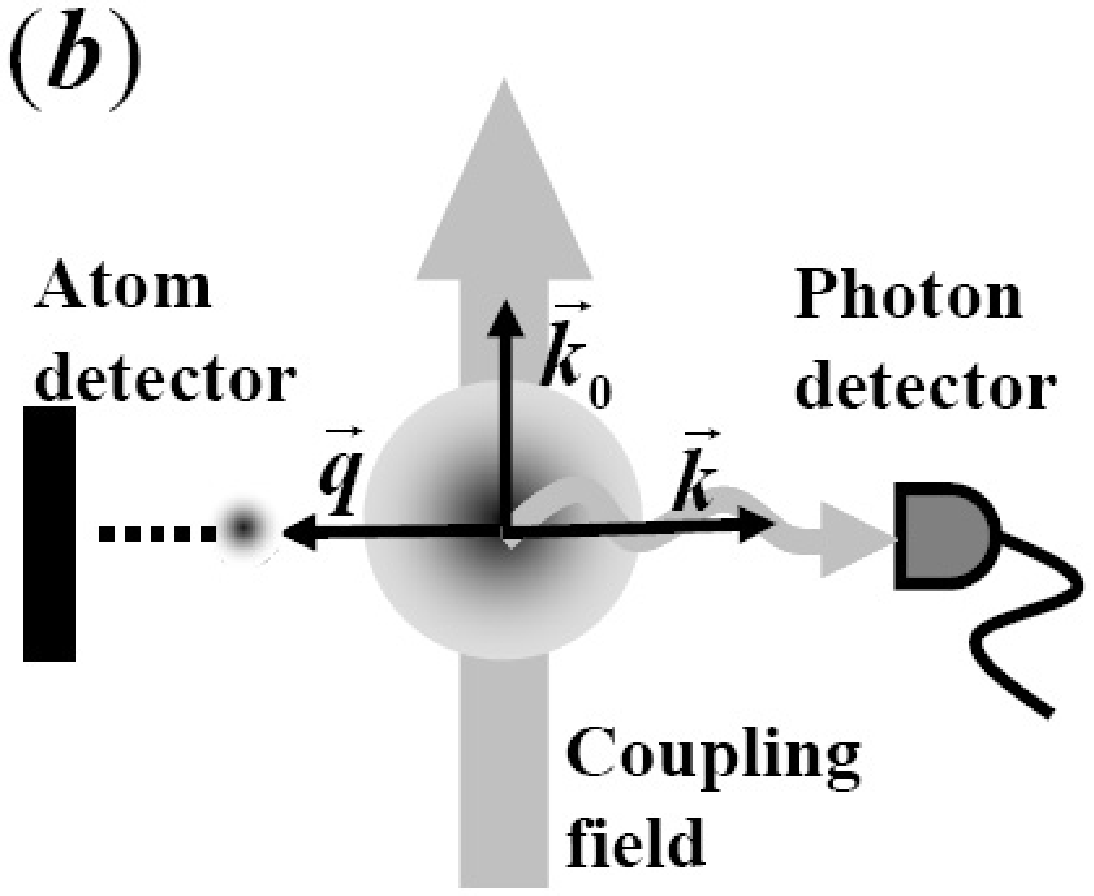}
\caption{(a) Atomic configuration with two well--separated upper
levels. The momentum entanglement can be controlled by the
classical coupling field through the SGC. (b) Schematic diagram
for the momentum detections. The detectors for the atom and photon
are restricted in one--dimension which is perpendicular to the
propagation direction of the coupling light.}
\end{figure}

Concerning the kinetic degrees of freedom, the Hamiltonian for the
system depicted in Fig. 1 (a) can be written with the rotating
wave approximation (RWA) as:
\begin{eqnarray}
&&\hat{H}=\frac{(\hbar
\hat{\vec{p}})^{2}}{2m}+\sum_{\vec{k}}\hbar\omega_{\vec{k}}\hat{a}^{\dag}_{\vec{k}}\hat{a}_{\vec{k}}
+\sum_{j=1}^{2}\hbar\omega_{jc}\hat{\sigma}_{jj}+\hbar\omega_{bc}\hat{\sigma}_{bb}\nonumber\\
&&+\hbar\sum_{\vec{k}}\left[
g_{1}(\vec{k})\hat{\sigma}_{c1}\hat{a}^{\dag}_{\vec{k}}e^{-i\vec{k}\cdot\vec{r}}
+g_{2}(\vec{k})\hat{\sigma}_{c2}\hat{a}^{\dag}_{\vec{k}}e^{-i\vec{k}\cdot\vec{r}}+{\rm
H.C.}\right]\nonumber\\
&&+\hbar \left(
\Omega_{1}e^{-i\nu_{0}t}e^{i\vec{k}_{0}\cdot\vec{r}}\hat{\sigma}_{1b}+
\Omega_{2}e^{-i\nu_{0}t}e^{i\vec{k}_{0}\cdot\vec{r}}\hat{\sigma}_{2b}+{\rm
H.C.} \right),\nonumber \\
\end{eqnarray}
where $\hbar \hat{\vec{p}}$ and $\vec{r}$ denote the atomic
center--of--mass momentum and position operators.
$\hat{\sigma}_{\alpha\beta}$ denotes the atomic operator
$|\alpha\rangle \langle \beta|$ ($\alpha,\beta = 1,2,b,c$), and
$\hat{a}_{\vec{k}}$ ($\hat{a}^{\dag}_{\vec{k}}$) is the
annihilation (creation) operator for the $k$th photonic mode with
wave vector $\vec{k}$ and frequency $\omega_{\vec{k}}=ck$, where
we use $\vec{k}$ to include both the momentum and polarization of
the photonic mode for simplicity. $g_{j}(\vec{k})$ is the coupling
coefficient for the transition $|j\rangle\rightarrow|c\rangle$
$(j=1,2)$ and $\Omega_{j}$ denotes the Rabi frequency for the
coupling $|j\rangle \leftrightarrow |b\rangle$ $(j=1,2)$. For the
convenience in further calculations, we may neglect the dependence
on $\vec{k}$ for $g_{j}(\vec{k})$ and treat them as constants.
$\nu_{0}$ and $\vec{k}_{0}$ denotes the frequency and the wave
vector of the coupling light, and $\omega_{\alpha\beta}$ is used
to represent the frequency difference as:
$\omega_{\alpha\beta}\equiv\omega_{\alpha}-\omega_{\beta}$. As the
evolution is considered in a close system, it is convenient to
expand the photon--atom state in the Schr\"{o}dinger picture as:
\begin{eqnarray}
|\psi\rangle&=&\sum_{\vec{q}}\left[ a_{1}(\vec{q})|\vec{q},0,1
\rangle+a_{2}(\vec{q})|\vec{q},0,2
\rangle+b(\vec{q})|\vec{q},0,b\rangle \right]\nonumber\\
&+&\sum_{\vec{q},\vec{k}}c(\vec{q},\vec{k})|\vec{q},\vec{k},c
\rangle,
\end{eqnarray}
where the arguments in the kets denote, respectively, the wave
vector of the atom, the photon, and the atomic internal state.

From the Schr\"{o}dinger equation one may yield the dynamic
equations for the system with the Born--Markov approximation. With
the transformation to the slowly varying parts:
\begin{eqnarray}
a_{1}(\vec{q})&=&e^{-i[T(\vec{q})+\omega_{1c}]t}\cdot
A_{1}(\vec{q}),\\
a_{2}(\vec{q})&=&e^{-i[T(\vec{q})+\omega_{2c}]t}\cdot
A_{2}(\vec{q}),\\
b(\vec{q}-\vec{k}_{0})&=&e^{-i[T(\vec{q}-\vec{k}_{0})+\omega_{bc}]t}\cdot
B(\vec{q}-\vec{k}_{0}),\\
c(\vec{q},\vec{k})&=&e^{-i[T(\vec{q})+\omega_{\vec{k}}]t}\cdot
C(\vec{q},\vec{k}),
\end{eqnarray}
where $T(\vec{p})\equiv\hbar\vec{p}^{2}/2m$ and
$\Delta_{j}\equiv\omega_{jb}-\nu_{0}$ $(j=1,2)$, we yield:
\begin{eqnarray}
i\frac{{\rm d}A_{1}(\vec{q})}{{\rm d}t}&=&\Omega_{1}e^{i\Delta_{1}t}B(\vec{q}-\vec{k}_{0})-\frac{i\gamma_{1}}{2}A_{1}(\vec{q})\nonumber\\
&-&\frac{i \epsilon\sqrt{\gamma_{1}\gamma_{2}}}{2}A_{2}(\vec{q})e^{i\omega_{12}t},\\
i\frac{{\rm d}A_{2}(\vec{q})}{{\rm d}t}&=&\Omega_{2}e^{i\Delta_{2}t}B(\vec{q}-\vec{k}_{0})-\frac{i\gamma_{2}}{2}A_{2}(\vec{q})\nonumber\\
&-&\frac{i \epsilon\sqrt{\gamma_{1}\gamma_{2}}}{2}A_{1}(\vec{q})e^{-i\omega_{12}t},\\
i\frac{{\rm d}B(\vec{q}-\vec{k}_{0})}{{\rm d}t}&=&\Omega_{1}^{*}e^{-i\Delta_{1}t}A_{1}(\vec{q})+\Omega_{2}^{*}e^{-i\Delta_{2}t}A_{2}(\vec{q}),\\
i\frac{{\rm d}C(\vec{q},\vec{k})}{{\rm
d}t}&=&g_{1}e^{i[T(\vec{q})-T(\vec{q}+\vec{k})+\omega_{\vec{k}}-\omega_{1c}]t}\cdot
A_{1}(\vec{q}+\vec{k})\nonumber\\
&+&g_{2}e^{i[T(\vec{q})-T(\vec{q}+\vec{k})+\omega_{\vec{k}}-\omega_{2c}]t}\cdot
A_{2}(\vec{q}+\vec{k}),\nonumber\\
\end{eqnarray}
where $\gamma_{1,2}$ denote the linewidthes for the two upper
levels $|1\rangle$ and $|2\rangle$; and $\epsilon\equiv
\vec{\mu}_{1}\cdot\vec{\mu}_{2}/|\vec{\mu}_{1}|\cdot|\vec{\mu}_{2}|$
with $\vec{\mu}_{j}$ being the dipole moment for the transition
$|j\rangle\rightarrow|c\rangle$ $(j=1,2)$. As in the experiments
\cite{exp}, we restrict the detections for the photon and atom in
one dimension, which is also perpendicular to the propagation of
the coupling field, as depicted in Fig. 1 (b).

In order to have strong interference between the transitions
$|1\rangle\rightarrow|c\rangle$ and
$|2\rangle\rightarrow|c\rangle$, the dipoles for the transitions
should be parallel or antiparallel, i.e., $\epsilon=\pm 1$
\cite{SGC theory, SGC-induced entang}; furthermore, the dressed
states produced by the coupling field should be nearly degenerate,
which may be fulfilled when
$\Delta_{1}/\Delta_{2}\approx-\gamma_{1}/\gamma_{2}$ \cite{SGC
theory}. To be consistent with these restrictions, as in some
experiments \cite{SGC exp}, we assume $\epsilon=1$,
$\gamma_{1}=\gamma_{2}=\gamma$, $\Omega_{1}=\Omega_{2}=\Omega$;
and $\Delta_{2}=-\Delta$, $\Delta_{1}=(1+\delta)\Delta$, where
$\delta$ is a dimensionless small term controlled by the detuning
of the coupling field. The atom is initially prepared in state
$|b\rangle$ with momentum wavefunction as $G(\vec{q})\propto
\exp{[-(\vec{q}/\delta_{p})^{2}]}$, where $\delta_{p}$ denotes its
momentum variance.

With the above simplifications, from Eqs. (7) to (10), it is
straightforward to yield the solutions for the whole system. As a
result, the steady state atom--photon entangled wave function
reads:
\begin{eqnarray}
C(q,k,t\rightarrow\infty)\approx\chi_{0}\frac{\exp{[-(\Delta
q/\eta)^{2}]}}{-\lambda_{1}/\gamma+i(\Delta q+\Delta k)}.
\end{eqnarray}
The $\Delta q$, $\Delta k$ $\eta$ and $\lambda_{1}$ are defined in
the appendix, where the other related mathematical details are
also given. $\chi_{0}$ is a normalization factor.

\begin{figure}
\centering
\includegraphics[height=3cm]{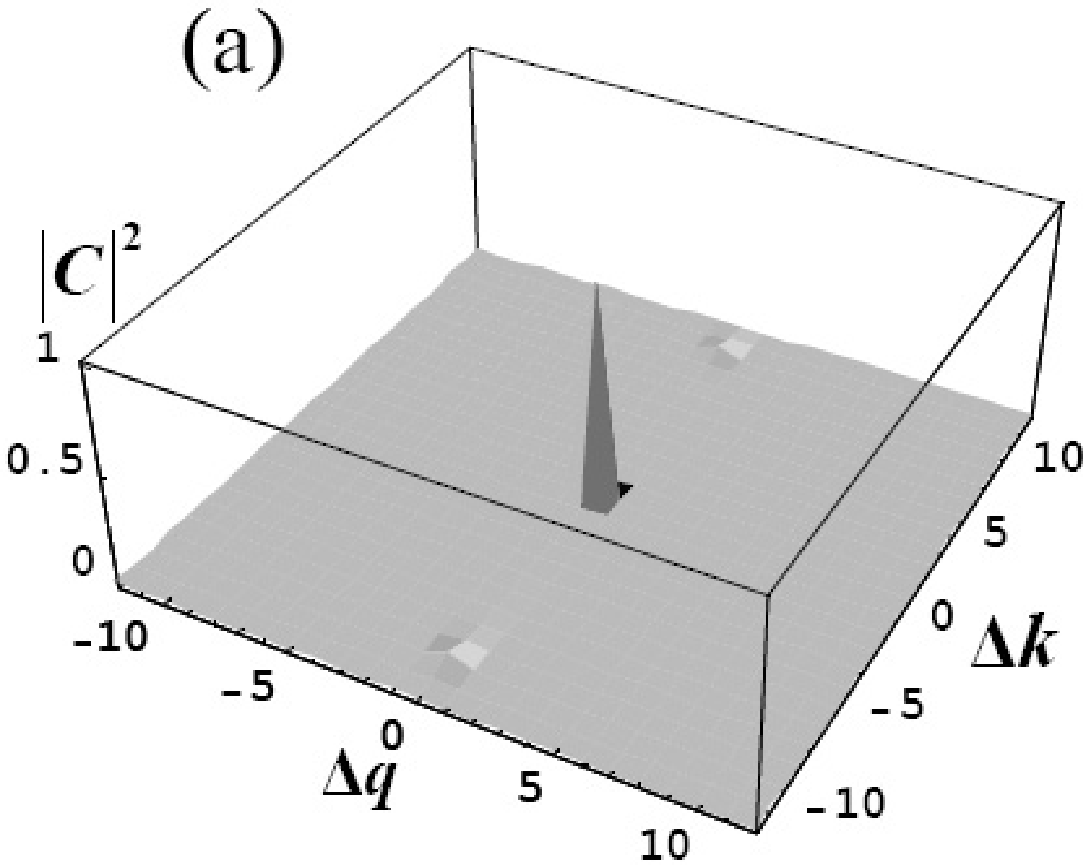}\ \ \includegraphics[height=3cm]{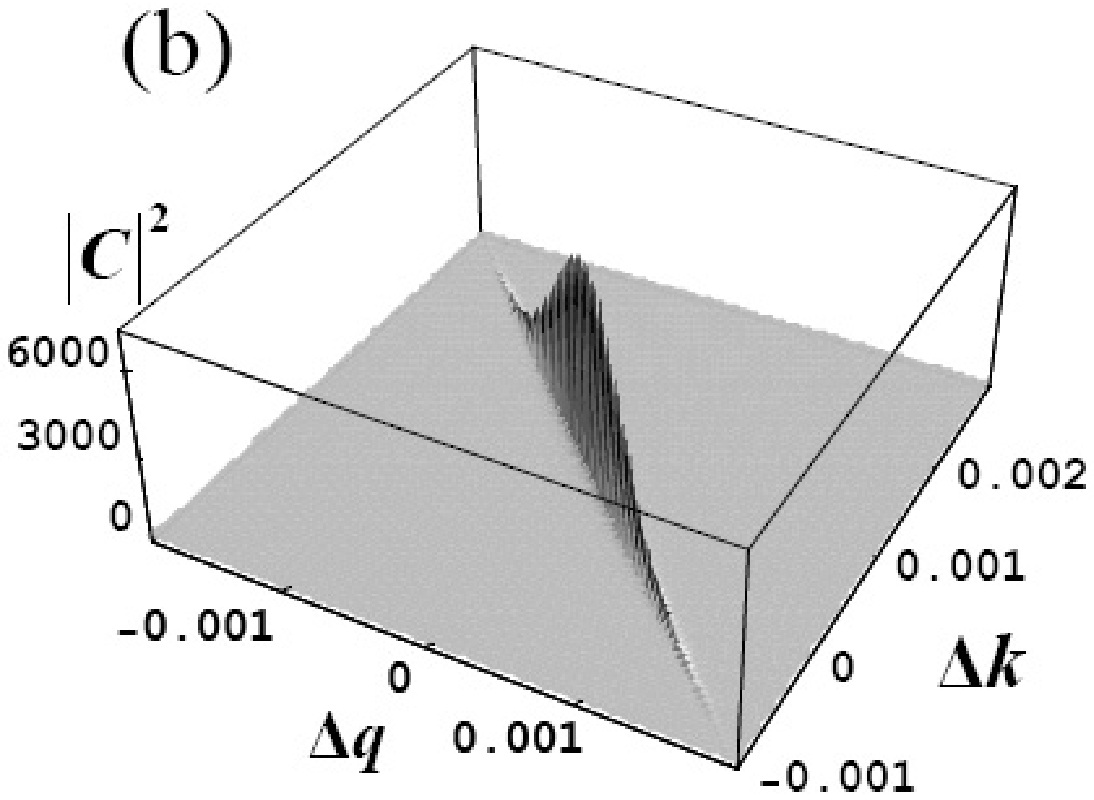}
\includegraphics[height=3cm]{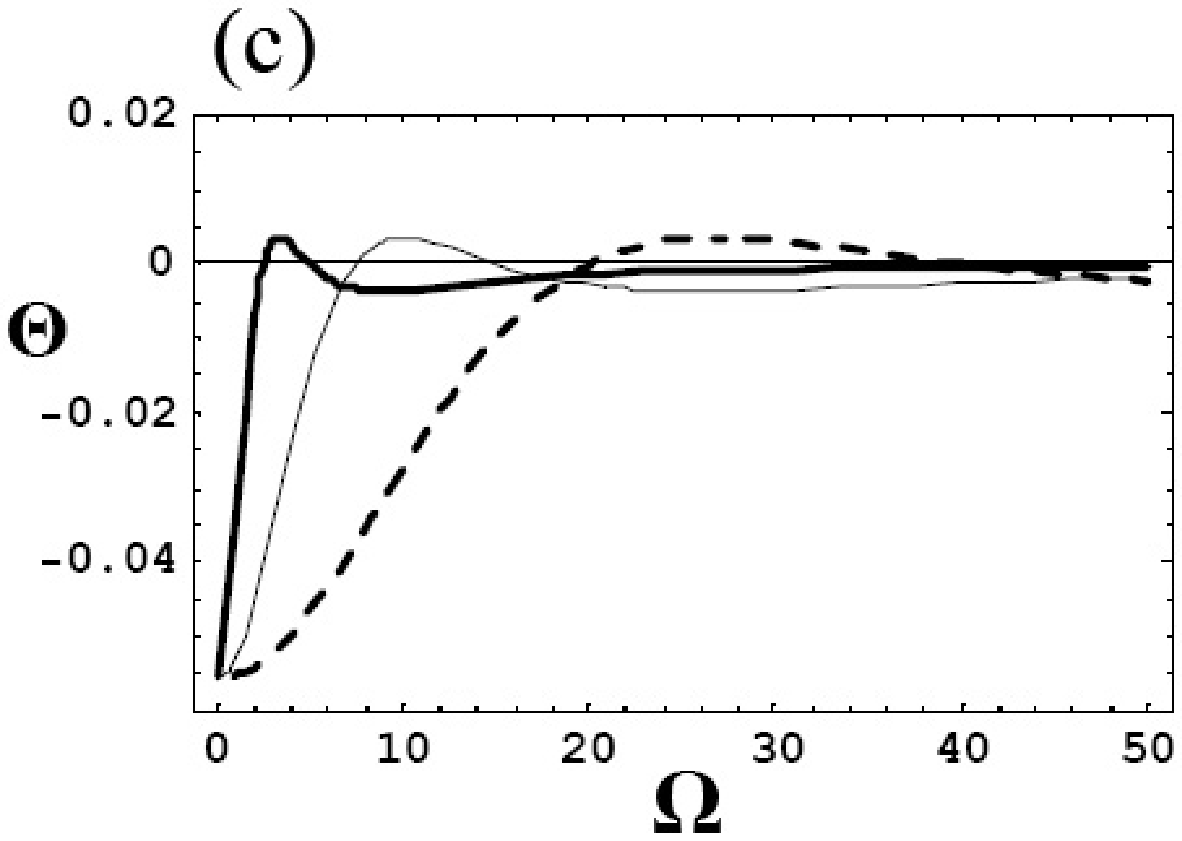}\ \includegraphics[height=3cm]{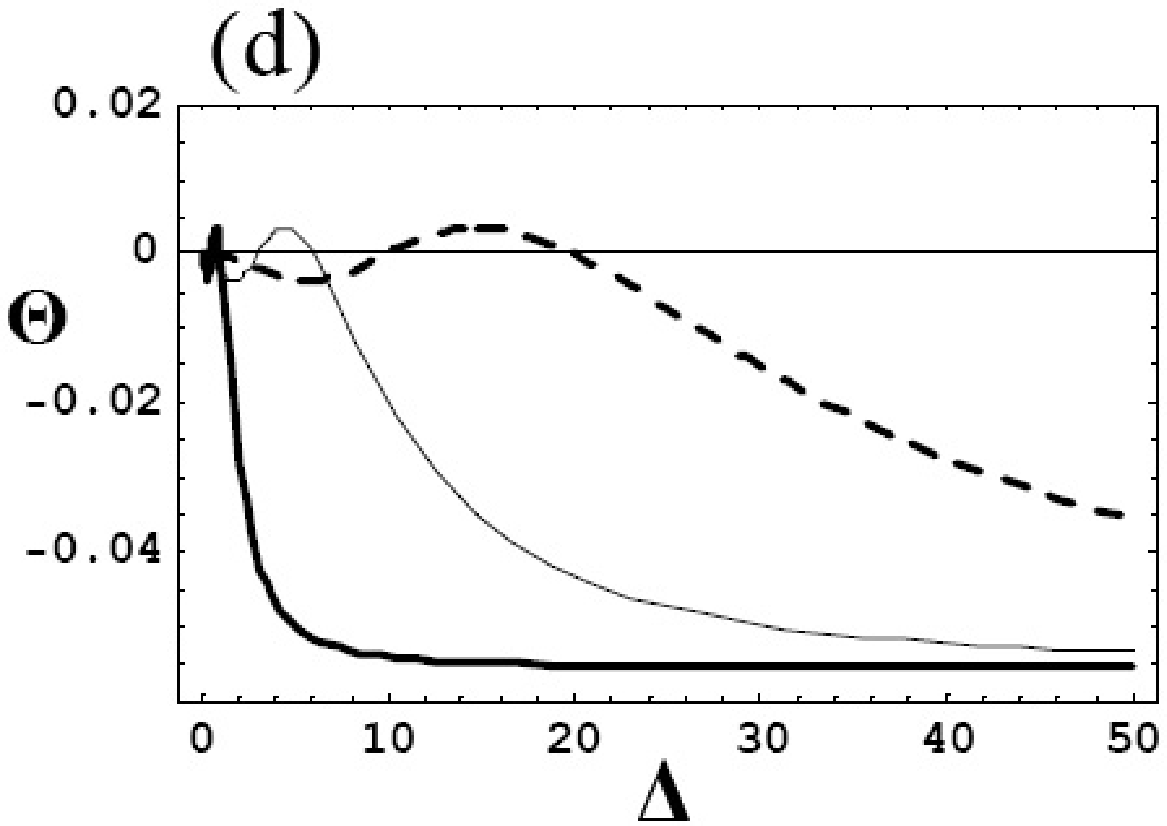}
\caption{(a) Distribution of $|C(\Delta q,\Delta k)|^{2}$ with
$\gamma=1$, $\Delta=10$, $\Omega=1$, $\delta=0.01$, $\eta$=0.001.
One sees that the central peak dominates the whole function. (b)
Plot of the local amplification of (a), which shows that $\Delta
q$ and $\Delta k$ are highly correlated in the central peak. (c)
Plot of the function $\Theta(\Omega)$ with $\Delta$ specified as
5, 15, 40, for the bold, thin and dashed lines, respectively. (d)
Plot of $\Theta(\Delta)$ with $\Omega$ specified as 0.5, 3, 10,
for the bold, thin and dashed lines, respectively. $\gamma=1$.}
\end{figure}

Theoretically, the entanglement of a pure state bipartite system
 can be completely evaluated by the Schmidt number $K$
\cite{Schmidt num}, which is defined as an estimation of the
number of modes that make up the Schmidt decomposition. From Eq.
(11), it is found that \cite{Singlephoton, scattering}
\begin{eqnarray}
K&&\approx 1+0.28\left[\frac{\eta}{|{\rm
Re}(\lambda_{1}/\gamma)|}-1\right], \nonumber \\
&&\approx \frac{0.28
\eta}{|\Theta(\Delta/\gamma,\Omega/\gamma)|\delta^{2}},
\end{eqnarray}
where the function $\Theta(\Delta/\gamma,\Omega/\gamma)$ is
defined in the appendix and depicted in Figs. 2 (c) and (d). With
respect to the realistic detections in experiment \cite{exp},
however, the degree of entanglement is better to be characterized
by the ratio ($R$) of the unconditional and conditional variances
in the momentum detections \cite{3-D spontaneous, GR,
photoionization}, i.e.:
\begin{eqnarray}
&&R\equiv \delta q^{{\rm single}}/\delta q^{{\rm coin}},\\
&&\delta q^{{\rm single}}\equiv \langle\hat{p}^{2}\rangle_{{\rm
single}}-\langle\hat{p}\rangle_{{\rm single}}^{2},\\
&&\delta q^{{\rm coin}}\equiv \langle\hat{p}^{2}\rangle_{{\rm
coin}}-\langle\hat{p}\rangle_{{\rm coin}}^{2},
\end{eqnarray}
where $\delta q^{{\rm single}}$ is defined as the variance of the
atomic momentum with single--particle detection, and $\delta
q^{{\rm coin}}$ denotes the variance obtained by the coincidence
detection on both the atom and the photon. From Eqs. (11) and
(13), we yield:
\begin{eqnarray}
R\approx\frac{\eta}{1.6|{\rm
Re}(\lambda_{1}/\gamma)|}=\frac{\eta}{1.6|\Theta
(\Delta/\gamma,\Omega/\gamma)|\delta^{2}}\approx 2.2K,
\end{eqnarray}
where one sees that the Schmidt number $K$ can therefore be
obtained experimentally by measuring the $R$ ratio \cite{exp}.

\begin{figure}
\centering
\includegraphics[height=3.7cm]{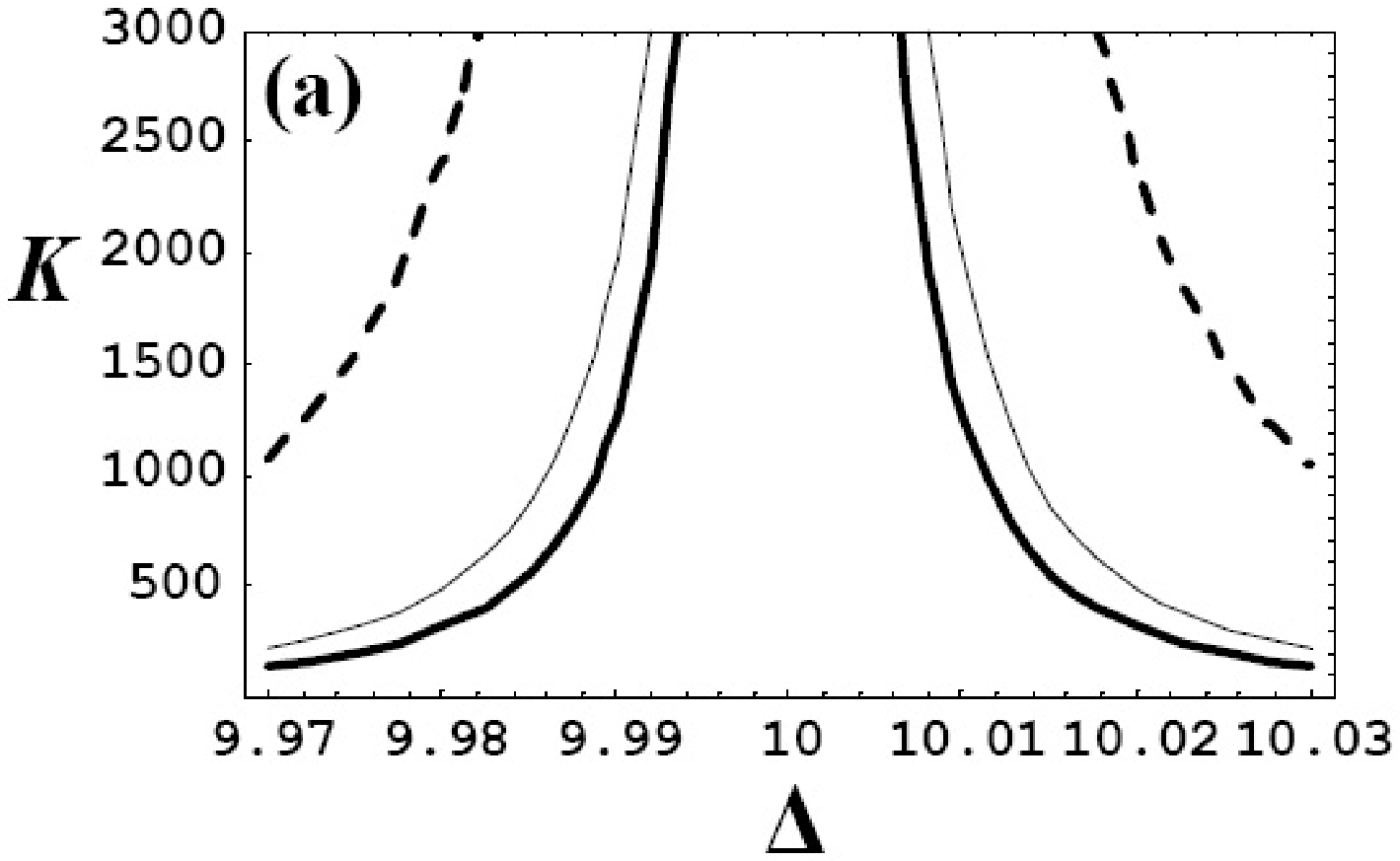}\ \ \ \includegraphics[height=3.5cm]{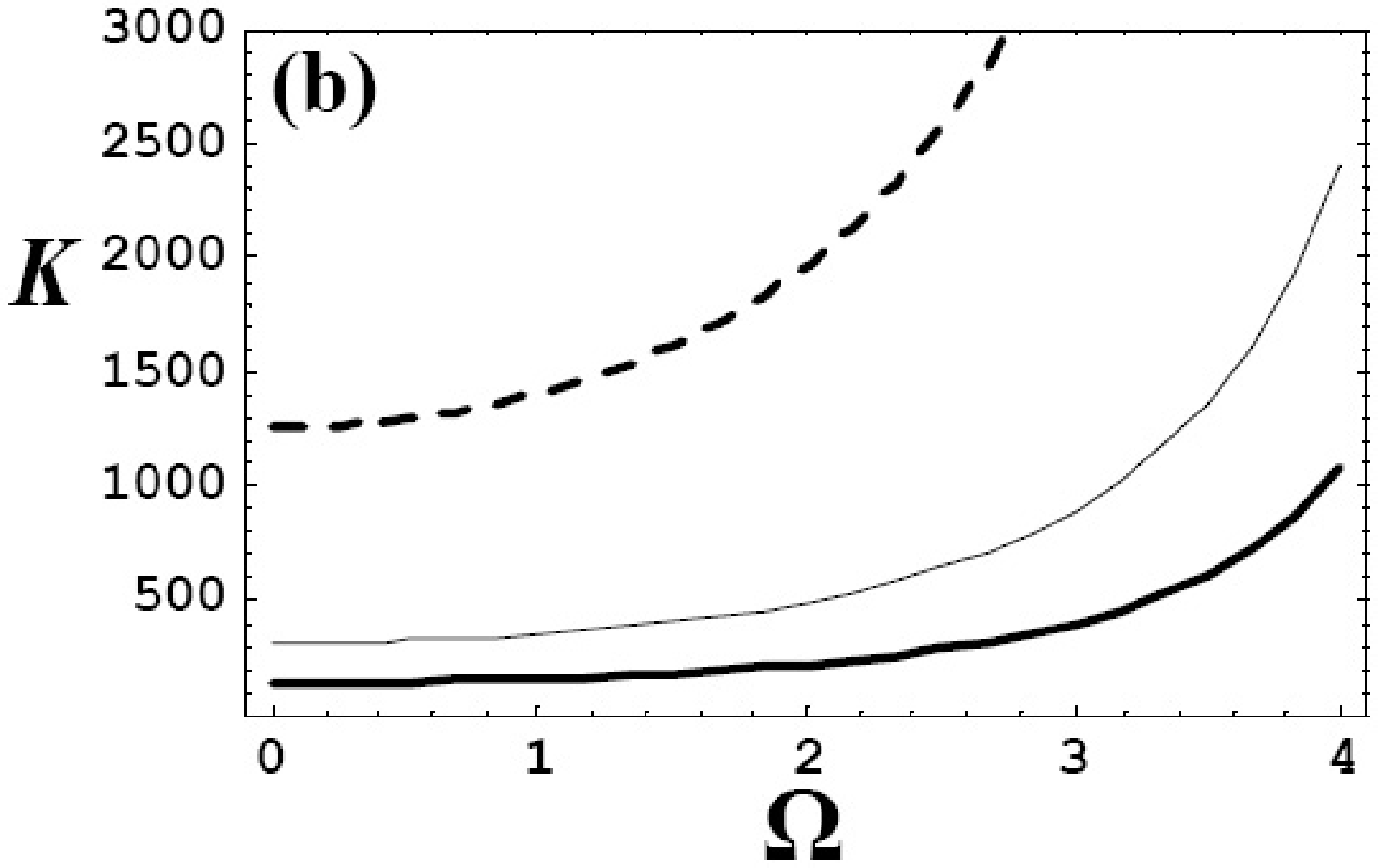}
\caption{Relations between the Schmidt number $K$ and the coupling
field with $\gamma=1$, $\omega_{12}=20$, $\eta=0.001$. (a) The
bold, thin and dashed lines are plotted with $\Omega=0.1$,
$\Omega=2$ and $\Omega=4$, respectively. (b) The bold, thin and
dashed lines are plotted with $\Delta=9.97$, $\Delta=9.98$ and
$\Delta=9.99$, respectively.}
\end{figure}

With Eq. (12) or (16), one sees that the entanglement is most
sensitive with the relative detuning $\delta$, therefore with the
frequency of the coupling field. With the relation
$\omega_{12}=\Delta_{1}-\Delta_{2}=(2+\delta)\Delta$, we plot in
Fig. 3 the degree of entanglement $K$ with respect to the detuning
$\Delta=\nu_{0}-\omega_{2b}$ and Rabi frequency $\Omega$ of the
coupling field. From Fig. 3 and Eq. (12), one sees that the
entanglement can be effectively enhanced by controlling the
coupling field, and when $\nu_{0}\rightarrow
\omega_{2b}+\omega_{12}/2$, we have $K\rightarrow \infty$.

Different to some previous works \cite{scattering,GR}, the
superhigh entanglement produced with this scheme is due to the
atomic spontaneously generated coherence (SGC) \cite{SGC
theory,SGC-induced entang}. With parallel dipoles ($\epsilon=1$),
the photon emitted along the two transition pathes
$|1\rangle\rightarrow|c\rangle$ and
$|2\rangle\rightarrow|c\rangle$ will interfere and modify the
momentum entanglement with the recoiled atom as a result. In
recent studies, it is found that, with nearly degenerate upper
levels and proper atomic coherence, the atom with SGC may exhibit
anomalous enhancement of momentum entanglement in the spontaneous
emission process \cite{SGC-induced entang}. In this scheme,
however, it is shown that, even with well--separated upper levels
and realistic conditions \cite{SGC exp}, the entanglement could
also be highly increased with the quantum interference which can
be controlled by a classical light field. Therefore, the proposed
scheme can most probably be used to produce highly entangled
atom--photon pairs in realistic applications.

\section{disentanglement}

As in the preceding section, to study the generation of momentum
entanglement \cite{Singlephoton,scattering,GR,3-D spontaneous}, it
is usually convenient to assume the entangled system to be a close
pure state system. However, in a realistic environment with $T\gg
0{\rm K}$, the interaction with environment will make the
entangled system into a mixed state, and, as a result, cause the
disentanglement. Actually, only when the disentangling process is
much slower than the generation of entanglement, the entangled
system can well be approximated by a pure state. With these
considerations, it is then possible to give out an upper bound for
the entanglement that could be produced reliably in the
environment.

Concerning the momentum entanglement, the disentanglement is
caused by the momenta exchange with the environment which may be
composed of background atoms and photons. Theoretically, the
influence from the background atoms can be eliminated by using a
high vacuum system; therefore, in order to study the unavoidable
disentanglement, we can simplify the environment as a heatbath of
background photons, coupled only to the entangled atom, as shown
in Fig. 4 (a). In order to give a general analysis for this
incoherent process, the atom is simplified as a two--level system
with resonant frequency $\omega_{a}$, then the Hamiltonian of the
total system under RWA is:
\begin{eqnarray}
\hat{H}_{{\rm tot}}&=& \hat{H}_{{\rm S}}+\hat{H}_{{\rm B}}+\hat{H}_{{\rm I}},\\
\hat{H}_{{\rm S}}&=& \frac{(\hbar
\hat{p})^{2}}{2m}+\hbar\omega_{a}\hat{\sigma}_{22}+\sum_{k}\hbar\omega_{k}\hat{b}^{\dag}_{k}\hat{b}_{k},\\
\hat{H}_{{\rm B}}&=&\sum_{k}\hbar\omega_{k}\hat{a}^{\dag}_{k}\hat{a}_{k}, \\
\hat{H}_{{\rm I}}&=&
\hbar\sum_{k}\left[g(k)\hat{\sigma}_{12}\hat{a}^{\dag}_{k}e^{-ikr}+{\rm
H.C.} \right],
\end{eqnarray}
where $\hat{H}_{{\rm S}}$, $\hat{H}_{{\rm B}}$ and $\hat{H}_{{\rm
I}}$ denote the Hamiltonians for the system (the entangled
atom--photon pair), the heatbath and the interaction between them,
respectively. $\hat{b}_{k}$ ($\hat{b}^{\dag}_{k}$) is the
annihilation (creation) operator for the entangled single--photon
in its $k$th mode, whereas $\hat{a}_{k}$ and $\hat{a}^{\dag}_{k}$
are those for the photons in the heatbath.

It is known that the density matrix of the total system
$\rho_{{\rm tot}}$ obeys the Liouville equation:
\begin{eqnarray}
\dot{\rho}_{\rm tot}=\mathcal{L}_{{\rm tot}}(\rho_{{\rm
tot}})=(\mathcal{L}_{{\rm S}}+\mathcal{L}_{{\rm
B}}+\mathcal{L}_{{\rm I}})\rho_{{\rm tot}},
\end{eqnarray}
where the superoperators $\mathcal{L}_{{\rm tot}}$,
$\mathcal{L}_{{\rm S}}$, $\mathcal{L}_{{\rm B}}$ and
$\mathcal{L}_{{\rm I}}$ are defined as: $ \mathcal{L}_{{\rm
tot}}(*)\equiv -\frac{i}{\hbar}[\hat{H}_{{\rm tot}},*]$, etc.. In
order to reveal the dynamic evolution for the entangled system, we
should adiabatically eliminate the heatbath terms from Eq. (21) to
obtain the master equation for the entangled system.

To proceed, we define the reduced density matrix for the system as
$\rho\equiv {\rm Tr_{B}}(\rho_{{\rm tot}})$ and a ``projection
state'' as $v\equiv {\rm Tr_{B}}(\rho_{{\rm
tot}})\otimes\rho_{{\rm B}}$, where the trace ``${\rm Tr_{B}}$''
is taken over the heatbath space and $\rho_{{\rm B}}$ denotes the
initial state of the heatbath. As the coupling is weak, from Eq.
(21), we yield the equation for the projection state as:
\begin{widetext}
\begin{eqnarray}
\dot{v}=\mathcal{L}_{{\rm S}}v -i \rho_{{\rm B}}\otimes {\rm
Tr_{B}}\left(\mathcal{L}_{{\rm I}}\int^{\infty}_{0}{\rm d}\tau
\left[\sum_{k}g(k)\hat{a}^{\dag}_{k}e^{-iw_{k}\tau}\Lambda(\tau)+{\rm
H.C.}\ ,\ \rho\otimes\rho_{{\rm B}}\right] \right) ,
\end{eqnarray}
\end{widetext}
where the Markov approximation and the nonrelativistic
approximation $\hbar k/mc \ll 1$ are used, and
$\Lambda(\tau)\equiv e^{-i\hat{H}_{{\rm
S}}\tau/\hbar}\hat{\sigma}_{12}e^{-ikr}e^{i\hat{H}_{{\rm
S}}\tau/\hbar}$.

From Eq. (22), we yield the master equation with the Lindblad form
\cite{Lindblad} as:
\begin{widetext}
\begin{eqnarray}
\dot{\rho}&=&-\frac{i}{\hbar}[\hat{H}_{{\rm
S}},\rho]-\Gamma_{a}\left[
D(\hat{\sigma}_{11}\rho+\rho\hat{\sigma}_{11})
+(1+D)(\hat{\sigma}_{22}\rho+\rho\hat{\sigma}_{22})\right]+(1+D)\Gamma_{a}\left[
\hat{\sigma}_{12}e^{-ik_{a}r}\rho\hat{\sigma}_{21}e^{ik_{a}r}+\hat{\sigma}_{12}e^{ik_{a}r}\rho\hat{\sigma}_{21}e^{-ik_{a}r}
\right]\nonumber \\
&+&D\Gamma_{a}\left[
\hat{\sigma}_{21}e^{ik_{a}r}\rho\hat{\sigma}_{12}e^{-ik_{a}r}+\hat{\sigma}_{21}e^{-ik_{a}r}\rho\hat{\sigma}_{12}e^{ik_{a}r}
\right],
\end{eqnarray}
\end{widetext}
where $D\equiv {\rm
Tr_{B}}(\hat{a}^{\dag}_{k_{a}}\hat{a}_{k_{a}}\rho_{{\rm B}})$ is
the average number of the resonant photons in the heatbath; the
atomic linewidth is given as $\Gamma_{a}\equiv
\sum_{k}2\pi|g(k)|^{2}\delta(\omega-\omega_{a})$. It is natural to
assume that the heatbath is initially in the thermal equilibrium,
i.e., $\rho_{{\rm B}}=e^{-\hat{H}_{{\rm B}}/k_{B}T}/{\rm
Tr_{B}}(e^{-\hat{H}_{{\rm B}}/k_{B}T})$, then we have
$D=1/(e^{\hbar\omega_{a}/k_{B}T}-1)$, where $T$ is the temperature
of the heatbath.

The master equation, Eq. (23), describes the process of
disentanglement, where the entangled system is now in a mixed
state due to the interaction with the heatbath. Theoretically, the
entanglement of a mixed bipartite system can be evaluated with the
``entanglement of formation''\cite{ent of formation}. However, in
order to base our analysis on direct experimental test \cite{exp},
we use the defined $R$ ratio [cf. Eq. (13)] as the evaluation of
the entanglement.

\begin{figure}
\centering
\includegraphics[height=3.3cm]{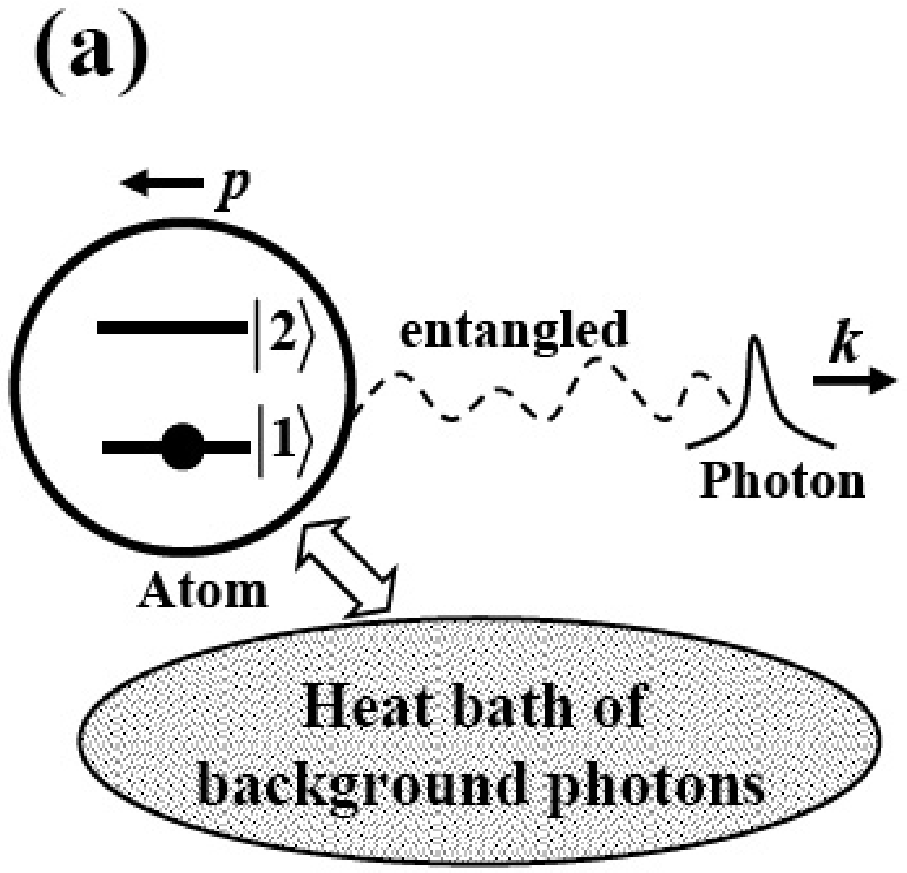}\ \ \includegraphics[height=3.5cm]{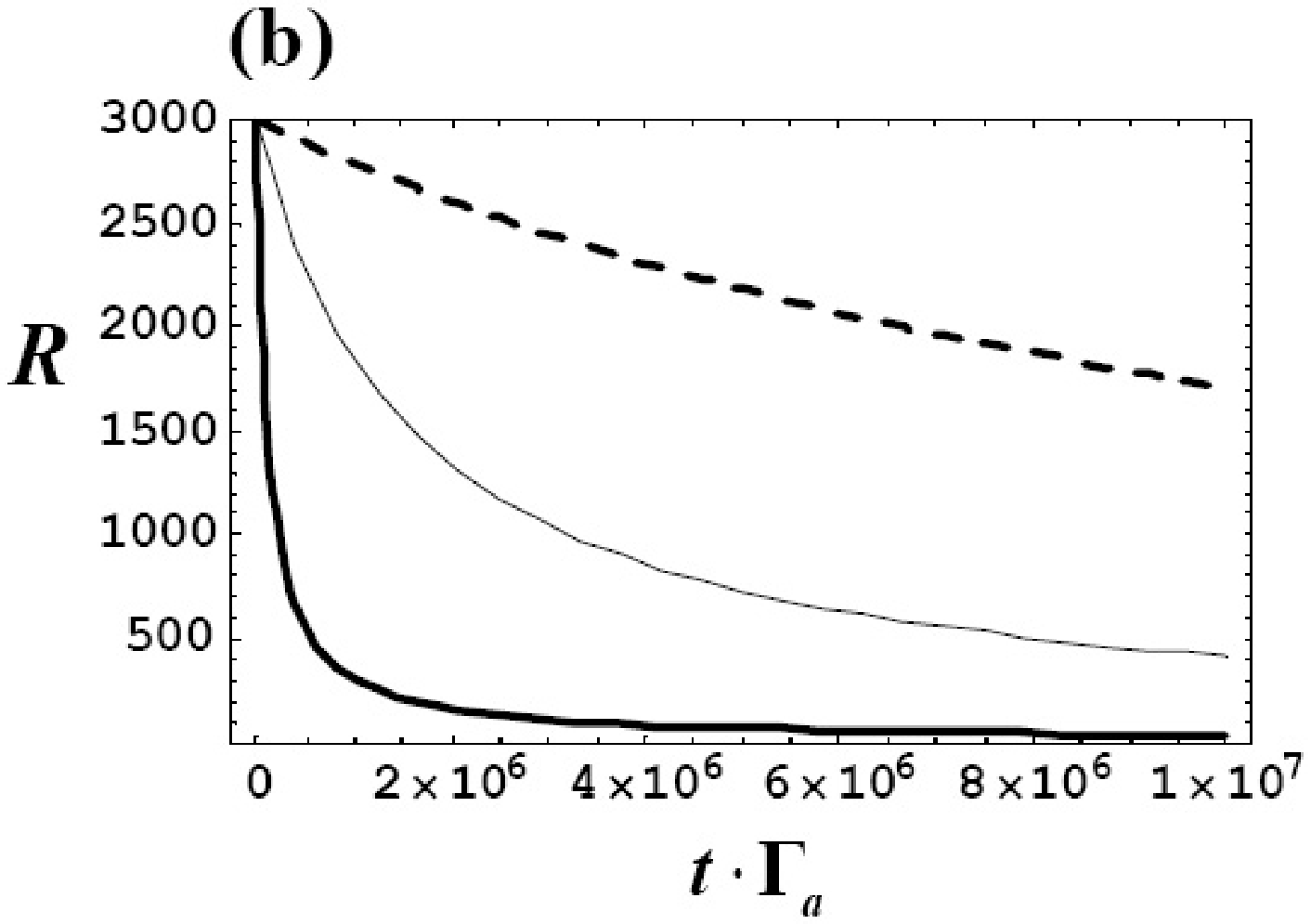}
\caption{(a) Schematic diagram of the disentanglement. The atom is
treated as a two--level system and the environment is simplified
as a heatbath of photons which is coupled to the atom through
momenta exchange. (b) $R(t)$ is plotted with $\delta q^{{\rm
single}}_{0}=16 k_{a}^{2}$ and $R_{0}$=3000. The temperature of
the environment is specified as $T=300{\rm\  K}$ for the bold
line, $T=270{\rm\  K}$ for the thin line and $T=250{\rm\  K}$ for
the dashed line.}
\end{figure}

As in Eqs. (14) and (15), the momentum variance of the
single--particle (the atom) measurement is calculated as:
\begin{eqnarray}
\delta q^{{\rm single}}={\rm Tr}(\hat{p}^{2}\rho)-[{\rm
Tr}(\hat{p}\rho)]^{2},
\end{eqnarray}
while that of the coincidence measurement is:
\begin{eqnarray}
\delta q^{{\rm coin}}&= &\frac{ \sum_{i=1}^{2}\int{\rm d}q \
q^{2}\rho(q,k_{0},i,q,k_{0},i)}{\sum_{i=1}^{2}\int{\rm d}q \
\rho(q,k_{0},i,q,k_{0},i)} \nonumber \\
 &-&\left(
\frac{\sum_{i=1}^{2}\int{\rm d}q \ q
\rho(q,k_{0},i,q,k_{0},i)}{\sum_{i=1}^{2}\int{\rm d}q \
\rho(q,k_{0},i,q,k_{0},i)}\right)^{2},
\end{eqnarray}
where $\rho(q,k,i,q',k',i')$ denotes the matrix element $\langle
q,k,i|\rho|q',k',i'\rangle$, and the photon is assumed to be
detected with some momentum $k_{0}$.

With Eqs. (23) to (25), it is straightforward to get:
\begin{eqnarray}
R(t)=R_{0}\cdot\frac{\delta q^{{\rm single}}_{0}/4D
k_{a}^{2}+\Gamma_{a} t}{\delta q^{{\rm single}}_{0}/4D
k_{a}^{2}+R_{0}\Gamma_{a} t},
\end{eqnarray}
where $R_{0}$ and $\delta q^{{\rm single}}_{0}$ are defined as
their initial values, $R_{0}\equiv R(t=0)$ and $\delta q^{{\rm
single}}_{0}\equiv\delta q^{{\rm single}}(t=0)$. It can be seen
from Fig. 4 (b) that, $R(t)$ decreases monotonously with time,
therefore, the characteristic time scale $\Delta t_{{\rm dis}}$
for the disentanglement can be defined as $R (t=\Delta t_{{\rm
dis}})=\frac{1}{2}R_{0} $. As $R_{0}\gg 1$, we yield:
\begin{eqnarray}
\Delta t_{{\rm dis}}\approx \frac{\delta q^{{\rm
single}}_{0}}{4DR_{0} k_{a}^{2}\Gamma_{a}}.
\end{eqnarray}

From Eq. (27), one sees that the ``disentangling time'' $\Delta
t_{{\rm dis}}$ is inversely proportional to the average number of
the resonant photons in the heatbath [$D(\omega_{a})$]. Therefore,
by decreasing the temperature of the environment it is possible to
significantly increase $\Delta t_{{\rm dis}}$ and then make the
entangled system quite robust in the environment. However, as the
temperature can never reach absolute zero, this kind of
disentanglement is ``unavoidable''. Furthermore, as in Eq. (27),
the disentangling time is also dependent on the initial
entanglement, i.e., $\Delta t_{{\rm dis}}\propto1/R_{0}$, which
indicates that, the better entangled system is more fragile in the
environment. Since $\Delta t_{{\rm dis}}\rightarrow 0$ when
$R_{0}\rightarrow \infty$, the ideal continuous EPR state
\cite{EPR} can never be reached in a realistic environment in this
sense.

The physical meaning for the dependence on $k_{a}$ and
$\Gamma_{a}$ in Eq. (27) is apparent: with larger energy and
shorter lifetime for the transitions, the environment will
exchange more momenta with the entangled atom per unit time, and
as a result, accelerate the disentanglement. When temperature is
low, the denominator in Eq. (27) has a sharp peak for the coupled
frequency $\omega_{a}$, which ensures the two--level approximation
reasonable for our treatment.

As stated at the beginning of this section, the entanglement
generated with Eqs. (12) and (16) is applicable only if the
disentangling time is much longer than the time scale for
producing the entanglement, i.e., $\Delta t_{{\rm dis}}\gg\Delta
t_{{\rm ent}}$. Therefore, with the Eqs. (27) and (A8), we yield
the inequality:
\begin{eqnarray}
R_{0}\ll 0.2\sqrt{\frac{\hbar
k_{1c}\delta_{p}^{3}}{Dmk_{a}^{2}\Gamma_{a}}} ,
\end{eqnarray}
which gives an upper bound for the entanglement that can be
produced in realistic environment with this scheme.

As in some reported experiments \cite{SGC exp}, the atomic
configuration with SGC as in Fig. 1 (a) can be realized by sodium
dimers. With the experimental conditions $\gamma \sim 10^{-7}
\omega_{1c}$, $\omega_{12}\sim 10\gamma$, $\Omega\sim \gamma$,
when the coupling field is tuned to $\Delta\sim \omega_{12}/2\sim
5\gamma$ with $\delta=10^{-2}$, from Eq. (A8), we have $\Delta
t_{{\rm ent}}\sim 1 {\rm ms}$. Take the time--of--flight into
account \cite{scattering}, the initial momentum variance can be
prepared as $\hbar \delta_{p}/m=1{\rm m/s}$, and then from Eq.
(16) we obtain a superhigh degree of entanglement as $R\approx
4600$ and $K\approx 2100$. To consider the disentanglement, we
take $\omega_{a}=5\times 10^{14}{\rm Hz}$ and
$\Gamma_{a}=10^{7}{\rm Hz}$ for estimations. With the environment
temperature $T=150{\rm K}$, from Eq. (27), we have $D\sim
10^{-11}$ and $\Delta t_{{\rm dis}}\sim 10^{4} {\rm s}$. One sees
that the relation $\Delta t_{{\rm ent}}\ll \Delta t_{{\rm dis}}$
can be well fulfilled. Therefore, under these conditions, the
robust highly entangled atom--photon pairs can be steadily
produced in the environment. Actually, from Eq. (28), we have an
upper bound as $R\ll 10^{7}$, which implies a strong ability of
producing entanglement with this scheme. On the other hand, if the
environment is at a high temperature, e.g., $T=400{\rm \ K}$, the
disentanglement will be strongly enhanced and we now have $\Delta
t_{{\rm ent}}\sim \Delta t_{{\rm dis}}$. With direct detections
\cite{exp}, it is then possible to observe all these phenomena in
experiment.

\section{conclusion}

In this paper, we demonstrate a novel scheme to produce superhigh
momentum entanglement between a single atom and a single photon
with the atomic SGC \cite{SGC-induced entang}. Under certain
experimental conditions \cite{SGC exp}, we show that the
entanglement can be effectively controlled by the classical
coupling field and may be very robust against the disentanglement
due to the environment. As we analyze the two physical processes
separately and both from the first principle, most of our
conclusions can directly apply to the previous models \cite{3-D
spontaneous,GR,scattering,Singlephoton}.

To give a better upper bound than Eq. (28), one may consider the
generation of entanglement together with the disentanglement at
the same time, which is also necessary to analyze the system when
$\Delta t_{{\rm ent}}\sim \Delta t_{{\rm dis}}$ rigorously.
However, this method is more complicated to be generalized and
will not change our above conclusions qualitatively. We plan to
give the details of this
method elsewhere in a future work.\\

This work is supported by the National Natural Science Foundation
of China (Grant No. 10474004), and DAAD exchange program:
D/05/06972 Projektbezogener Personenaustausch mit China
(Germany/China Joint Research Program).

\section*{Appendix: steady solutions of equations (7) to (10)}

In order to obtain Eq. (11), we define a matrix $M$ as:
\begin{align}
M=\left(\begin{array}{ccc}
\frac{\gamma_{1}}{2}+i\Delta_{1},&\frac{\epsilon\sqrt{\gamma_{1}\gamma_{2}}}{2},&i\Omega_{1}\\
\frac{\epsilon\sqrt{\gamma_{1}\gamma_{2}}}{2},&\frac{\gamma_{2}}{2}+i\Delta_{2},&i\Omega_{2}\\
i\Omega_{1}^{*},&i\Omega_{2}^{*},&0
\end{array}\right)\tag{A1},
\end{align}
its eigenvalues and eigenvectors are denoted, respectively, as
$\lambda_{j}$ and $(\alpha_{j},\beta_{j},\zeta_{j})^{{\rm T}}$
with $j=1,2,3$. Then from Eqs. (7) to (10), the steady state
solution of the entangled wave function can be written as:
\begin{align}
& C(\vec{q},\vec{k},t\rightarrow\infty)= \tag{A2} \\
&
\sum_{j=1}^{3}\frac{i(g_{1}\alpha_{j}+g_{2}\beta_{j})p_{j}G(\vec{q}+\vec{k}-\vec{k}_{0})}
{-\lambda_{j}+i\left[
T(\vec{q})-T(\vec{q}+\vec{k})+\omega_{\vec{k}}-(\omega_{bc}+\nu_{0})
\right]},\nonumber
\end{align}
where $p_{j}$ is determined by the initial conditions, and when
the atom is initially in $|b\rangle$, the restrictions are:
\begin{align}
\sum_{j=1}^{3}p_{j}\alpha_{j}=0,\ \
\sum_{j=1}^{3}p_{j}\beta_{j}=0,\ \
\sum_{j=1}^{3}p_{j}\zeta_{j}=1.\tag{A3}
\end{align}
As the detection is restricted in one dimension, the solution can
be simplified as:
\begin{align}
C(q,k)&\propto\sum_{j=1}^{3}\frac{(g_{1}\alpha_{j}+g_{2}\beta_{j})p_{j}e^{-(\Delta
q/\eta)^{2}}} {-\lambda_{j}/\gamma_{2}+i( \Delta q+\Delta k )}, \tag{A4} \\
& \equiv \sum_{j=1}^{3}L_{j}(\Delta q,\Delta k),\tag{A5}
\end{align}
where the effective wave vectors are defined as:
\begin{align}
\Delta
q&\equiv\frac{\hbar(\omega_{bc}+\nu_{0})}{mc\gamma_{2}}(q-\frac{\omega_{bc}+\nu_{0}}{c}),\tag{A6}\\
\Delta
k&\equiv\frac{ck-(\omega_{bc}+\nu_{0})}{\gamma_{2}},\tag{A7}
\end{align}
and $\eta\equiv\hbar(\omega_{bc}+\nu_{0})\delta_{p}/m\gamma_{2}c$.
We use $L_{1,2,3}(\Delta q,\Delta k)$ here to denote the three
different terms that make up the summation in Eq. (A4). From Eq.
(A4), it can be proved that:
\begin{eqnarray}
\int {\rm d}\Delta q {\rm d}\Delta k |L_{2,3}(\Delta q, \Delta k
)|^{2}=o(\Omega^{2}/\gamma^{2}) \ \ {\rm when} \ \Omega\rightarrow
0,\nonumber
\end{eqnarray}
which indicates that the $L_{1}(\Delta q, \Delta k)$ dominates the
summation when the coupling field is weak, as shown in Figs. 2 (a)
and 2 (b); therefore, Eq. (A4) can be well approximated by the
single--peak function as in Eq. (11).

To give further analysis for the entanglement, from Eq. (A1), we
yield ${\rm
Re}(\lambda_{1}/\gamma)\approx\Theta(\Delta/\gamma,\Omega/\gamma)\delta^{2}$,
where the value of $\Theta(\Delta/\gamma,\Omega/\gamma)$ is of
order $0.1$ or smaller as shown in Figs. 2 (c) and (d). Moreover,
the time scale for producing the entanglement can be characterized
as: $ \Delta t_{{\rm ent}}=1/|{\rm Re}(\lambda_{1})|=1/\gamma
|\Theta|\delta^{2}$, with Eq. (16), it may be written as:
\begin{align}
\Delta t_{{\rm ent}}\approx\frac{1.6R}{\eta\gamma}. \tag{A8}
\end{align}

\end{document}